\documentclass{elsart}
\usepackage{natbib}
\usepackage{amsmath}
\usepackage{natbib}
\usepackage{epsfig}

\begin{document}
\runauthor{Bolton et al.}
\begin{frontmatter}
\title{Follow-up of the 9C survey: Initial results}

\author[cambridge]{Rosemary C. Bolton}
\author[cambridge]{Garret Cotter}
\author[caltech]{T.J. Pearson}
\author[cambridge]{Guy G. Pooley}
\author[caltech]{A.C.S. Readhead}
\author[cambridge]{Julia M. Riley}
\author[cambridge]{Elizabeth M. Waldram}

\address[cambridge]{Astrophysics, Cavendish Laboratory, Cambridge CB3 0HE, UK}
\address[caltech]{California Institute of Technology, Pasadena, CA 91125,
  USA}

\begin{abstract}

We present initial results from a follow-up of the 9C survey, complete to
25 mJy at 15 GHz, designed to assemble and investigate a sample of
young radiosources. We have made radio continuum maps of 111 sources
at frequencies spanning 1.4---43 GHz, and classified them
according to their radio size and spectral index between 1.4 and 4.8
GHz. We find that selection at 15 GHz is twice as efficient at picking
Gigahertz Peaked Spectrum (GPS) sources as selection at $\sim 5$
GHz. Optical follow-up has now begun; imaging of GPS and
compact steep spectrum sources suggests that a significant
fraction of the host galaxies have close companions or disturbed
morphologies.

\end{abstract}
\end{frontmatter}

\section{Introduction}
\label{outline}

It is now widely believed, as evidenced by many contributions in these
proceedings, that the onset of AGN activity in galaxies is closely
linked to vigorous star-formation, possibly in the presence of
mergers.  However, the precise mechanism that causes radio sources to
trigger remains the subject of lively debate---see,
e.g.,\cite{2000AAS...19710908P} versus \cite{1995AAS...186.1502W}.  To
advance our understanding of this issue it is clearly necessary to
assemble samples of radiosources at the earliest possible times in
their evolution.

The Gigahertz Peaked Spectrum (GPS) and Compact Steep Spectrum (CSS)
sources are now thought to be the progenitors of the larger, older
members of the population. These young sources, often morphologically
classified as Compact Symmetric Objects (CSOs), have indeed in a few
cases been observed growing at sub-relativistic speeds by VLBI
\cite{2000ApJ...541..112T}. Unfortunately it has been difficult to
obtain flux-limited samples of GPS and CSS sources because large-area surveys
at frequencies higher than 5 GHz have been unavailable. Now, with the
advent of cm-wavelength Cosmic Microwave Background (CMB) experiments,
such surveys are being carried out to implement foreground source
subtraction. The Cambridge Ryle Telscope (RT) is undertaking a survey
at 15 GHz, the 9C survey (7), of CMB fields observed by the Very Small
Array (VSA).  Using 9C we have selected 156 sources on the basis of
their 15 GHz flux alone; these have all been followed up at
frequencies from 1.4 GHz to 43 GHz. Radio frequency data reduction is
now complete for two of the three sample fields and we present results
for these 111 sources. This sub-sample is complete to 25 mJy and
covers a total area of 140 square degrees.

\section{Radio and optical observations and analysis}

The first three regions of 9C coincide with the VSA compact array
fields, at $ 00^h 20^m$ $ +30^{\circ}$, $ 09^h 40^m$ $ +32^{\circ}$
and $ 15^h 40^m$ $ +43^{\circ}$.  Our sample is complete to 25 mJy in
the $00^h$, the $15^h$ field and for one-quarter of the $09^h$
field. The remaining three-quarters of the $09^h$ field is complete to
60 mJy. Continuum snapshot maps were made at 1.4, 4.8, 22 and 43 GHz
with the VLA and at 15 GHz with the RT between 2001 January and 2002
January.
All observations for
each region were made simultaneously to avoid problems introduced by
source variability.
%


We have used the radio fluxes to classify each source according to the
spectral index, {$\alpha$} between 1.4 and 4.8 GHz (we take $ S
\propto \nu^{-\alpha}$). We define three spectral classes: steep
spectrum sources with $\alpha^{1.4}_{4.8} \ge 0.5$, flat spectrum
sources with $-0.1 \le \alpha^{1.4}_{4.8} < 0.5 $, and a third class
containing those sources with $\alpha^{1.4}_{4.8} < -0.1 $. The last
of these includes
the GPS sources displaying a peak in
flux at 4.8, 15 or 22 GHz, and those still rising at 43 GHz, which are
assumed to be extreme versions of the GPS sources. We also classify
objects as compact if they have a radio size $\le 2^{\prime\prime}$,
and as extended otherwise. Of the 111 sources  here, 55\% are
compact and 45\% are extended in the radio. The numbers of sources in
each class are summarised in Table \ref{population}.
\begin{table}[h]
\begin{center}
\begin{tabular}{|l|l|l|l|l|l|l|}                                      
\hline
  &  Compact sources &  Extended sources & Total \\    
\hline
Steep spectrum  & 19 &  37   &  56  \\ 
Flat spectrum          &    23  &11  &34  \\
Peaked spectrum          &   19   &2 & 21  \\	
\hline  
Total         &   61   &50& 111 \\
\hline                
\end{tabular}
\caption{Numbers of sources of different classes for the 9C subsample
presented here.}
\label{population}
\end{center}
\end{table}

\begin{figure}[h]
\centerline{
 \epsfig{figure=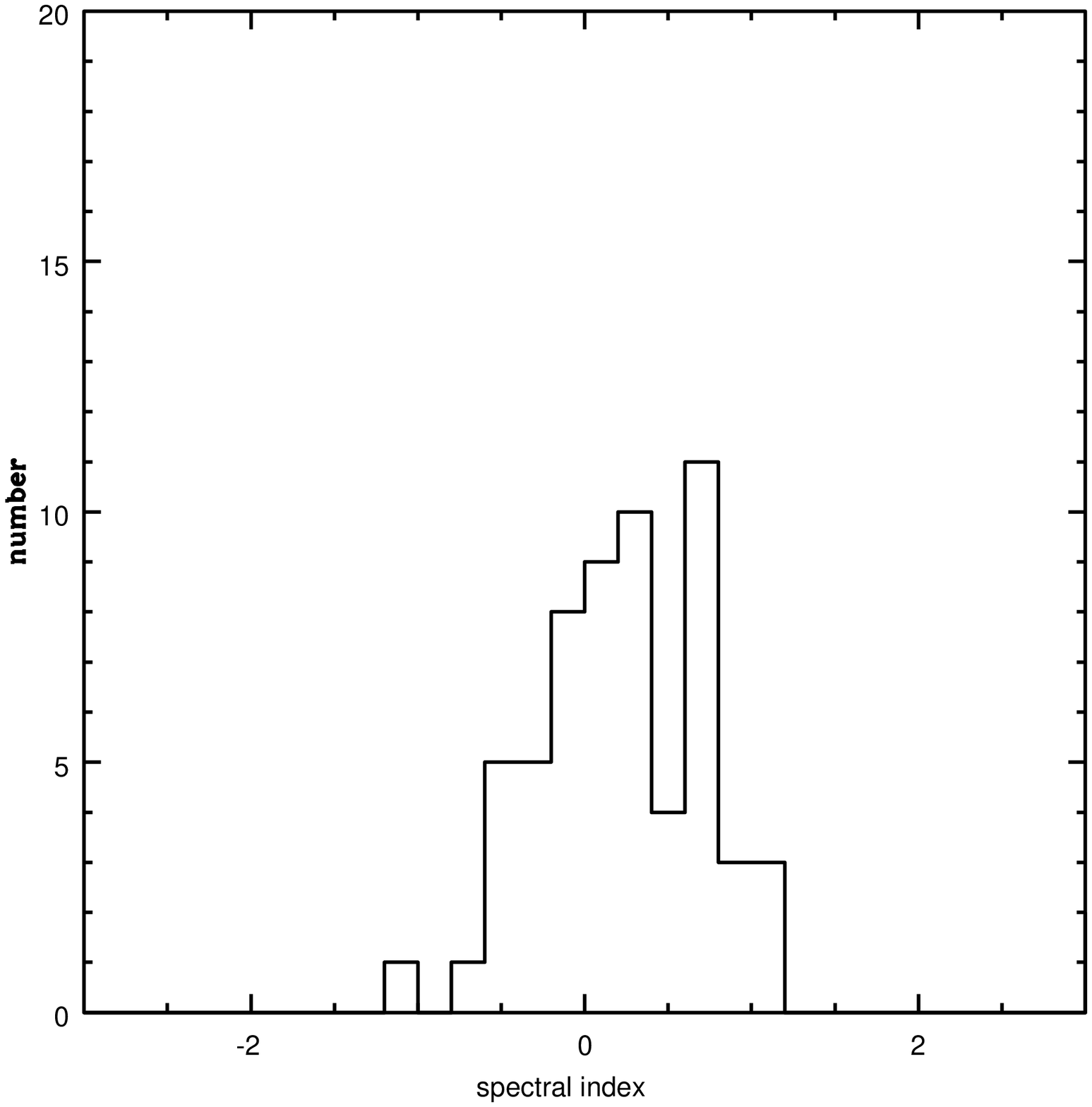,height=7cm,clip=}\qquad
  \epsfig{figure=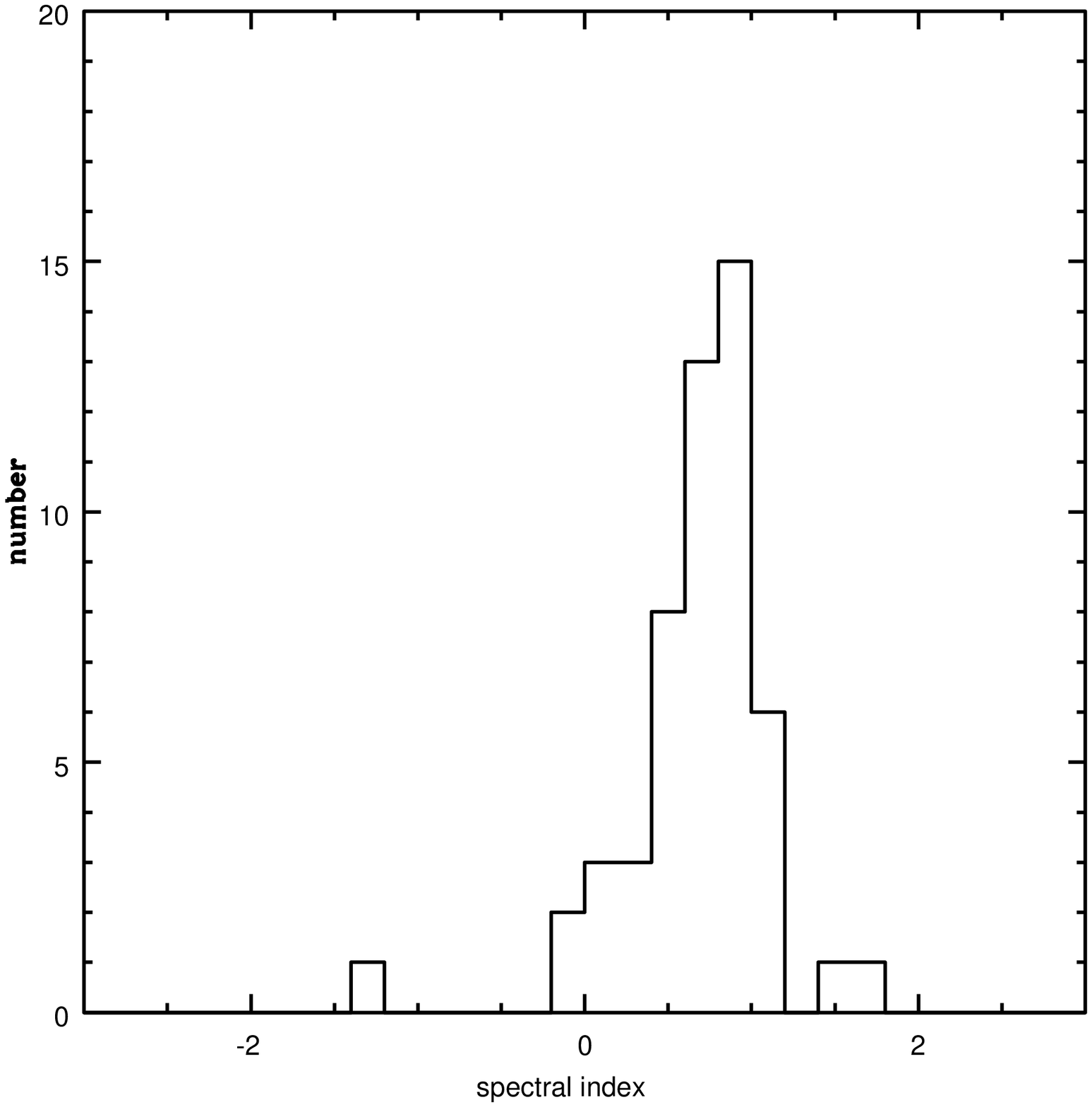,height=7cm,clip=}}
   \caption{Binned histograms of spectral index between 1.4 and 4.8
 GHz for the compact (left) and the extended (right) radio sources (right) (bin size 0.2).}
\label{size_alpha}
\end{figure}

It is already well established for samples selected at lower radio
frequencies that extended radio sources have steeper spectra than
compact sources (see, e.g. Peacock and Wall \cite{P1}, PW hereafter),
and the trend continues for this sample. Plotting the distribution of
spectral index for the two size classes (Figure \ref{size_alpha})
clearly shows this tendency: the median values of $\alpha^{1.4}_{4.8}$
are 0.21 and 0.76 for the compact and extended radio sources
respectively.
To examine how useful our selection process is as a means of
finding GPS sources we  compare our sample to PW, who studied a sample of bright
($S_{2.7}\geq1.5$ Jy) radio sources selected at 2.7 GHz. By combining
our flat and peaked sources into the same class (i.e $\alpha<0.5$) we
can directly compare our sample with  PW. The
percentages of sources falling into each class in the two samples are
shown in table \ref{PW_compare}. Our sample contains a lower fraction
of steep-spectrum sources than PW, but roughly the
same fraction of the compact sources  (note, though,
that PW calculate $\alpha$ between 2.7 and 5 GHz, and so will be more
likely to classify sources with standard convex spectra as
steep-spectrum).
\begin{table}[h]
\begin{center}
\begin{tabular}{|l|l|l|l|l|}                                      
\hline
  &Sample & Compact sources &  Extended sources& Total \\    
\hline
Steep spectrum  &This work& 17 &  33   &  50  \\ 
($\alpha \ge 0.5$)                &   PW    & 22 &  48   &  70  \\ 
\hline
``Flat'' spectrum   &This work &    38 &12  &50  \\
($\alpha < 0.5$)   &PW        &    27 & 3  &30  \\
\hline
Total    &This work     &   55   &45&  \\
Total     &   PW     &  49   & 51 &  \\
\hline                
\end{tabular}
\vspace{0.25cm}
\caption{Percentages of sources in different classes for this work and the PW sample.}
\label{PW_compare}
\end{center}
\end{table}
PW did not make any distinction between flat-spectrum and GPS sources,
but O'Dea \cite{O1} summarised the frequency of occurence of GPS and
compact steep spectrum (CSS) sources for flux-limited samples taken at
around 5 GHz. Using the values given by O'Dea and \cite{K1}, along
with the values from PW, the numbers of extended steep spectrum, CSS,
flat spectrum and GPS sources a sample selected at $\sim 5$ GHz should
contain can be estimated. These approximate percentage values, and
those of this work, are given in table \ref{5GHz_compare}. We find
roughly twice as many GPS sources as in samples selected at $\sim 5$ GHz,
but only about half as many CSS sources.
\begin{table}[ht]
\begin{center}
\begin{tabular}{|l|l|l|}                                      
\hline
 Selection Frequency & $\sim 5$ GHz & 15 GHz \\    
\hline
 Extended Steep Spectrum  & 40 &  33   \\ 
\hline
Compact Steep Spectrum  & 30 &   17  \\ 
\hline
Flat Spectrum   &   20  & 31 \\
\hline
Giga-Hertz Peaked Spectrum   &   10  & 19 \\
\hline                
\end{tabular}
\caption{Approximate percentages of different source classes for a typical $\sim 5$ GHz sample and this work.}
\label{5GHz_compare}
\end{center}
\end{table}

\begin{figure}[h]
\centering
\epsfig{figure=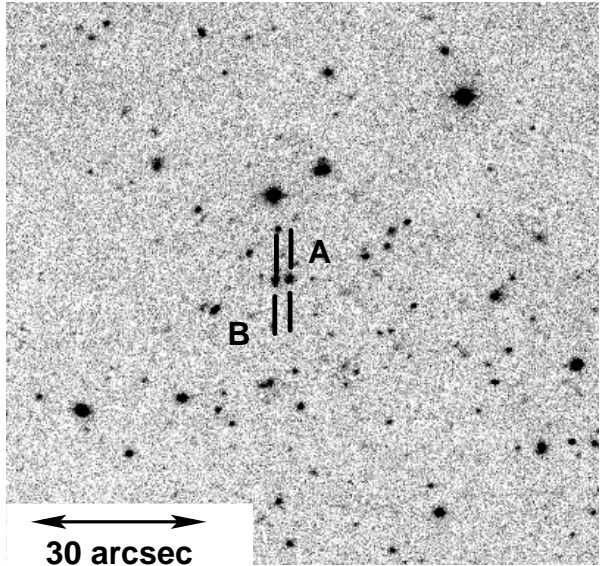,height=18cm,angle=270,clip=}

\caption{3600-s Palomar 60$^{\prime \prime}$ $I$-band image of a
$z=0.9$ system containing a 9C CSS galaxy. Both the radiogalaxy, A, and
its near companion, B, have disturbed morphologies and show bright AGN
emission lines. There is a clear excess of faint galaxies centred on
the radiosource; we speculate that this is a dynamically unrelaxed,
possibly merging, cluster.}
\label{cluster_image}
\end{figure}

A programme of optical imaging and spectroscopy is underway to obtain
identifications and redshifts for the compact sources. To date (2003
January) we have obtained imaging with the Palomar $ 60 ^{\prime
\prime}$ telescope of roughly two-thirds of the sample, including all
targets fainter than $R_{\rm AB} = 19.2$. While we caution that this
work is still incomplete, we note that the initial results
suggest that many of the compact sources have close companions or
disturbed morphology at optical wavelengths. We find that almost all
the low-$z$ host galaxies imaged to date have faint
companions within $\sim $ 10 arcsec. As we move to more distant host
galaxies, where
our imaging programme is thus far insufficiently deep to detect very faint
companions, several noticeably turbulent environments have been
discovered (Fig. \ref{cluster_image}). 

It is clearly now essential, in the light of the debate over major
versus minor interactions as the trigger for radiosource activity,
that we quantify these optical observations for the complete sample
and compare with control samples of both larger radiosources and
quiescent ellipticals.

\section{Discussion and summary}

In the standard picture of GPS sources, the energy density of the
synchrotron plasma in the lobes is sufficiently high that they remain
optically thick to higher radio frequencies \cite{K2}, with the
spectral peak indicating the frequency at which the lobe plasma
becomes optically thin. As the source grows to kpc size, the peak in
the spectrum remains, but falls to MHz frequencies and below, and a
CSS source is observed. 

\begin{figure}[h]
 \centerline{
   \epsfig{figure=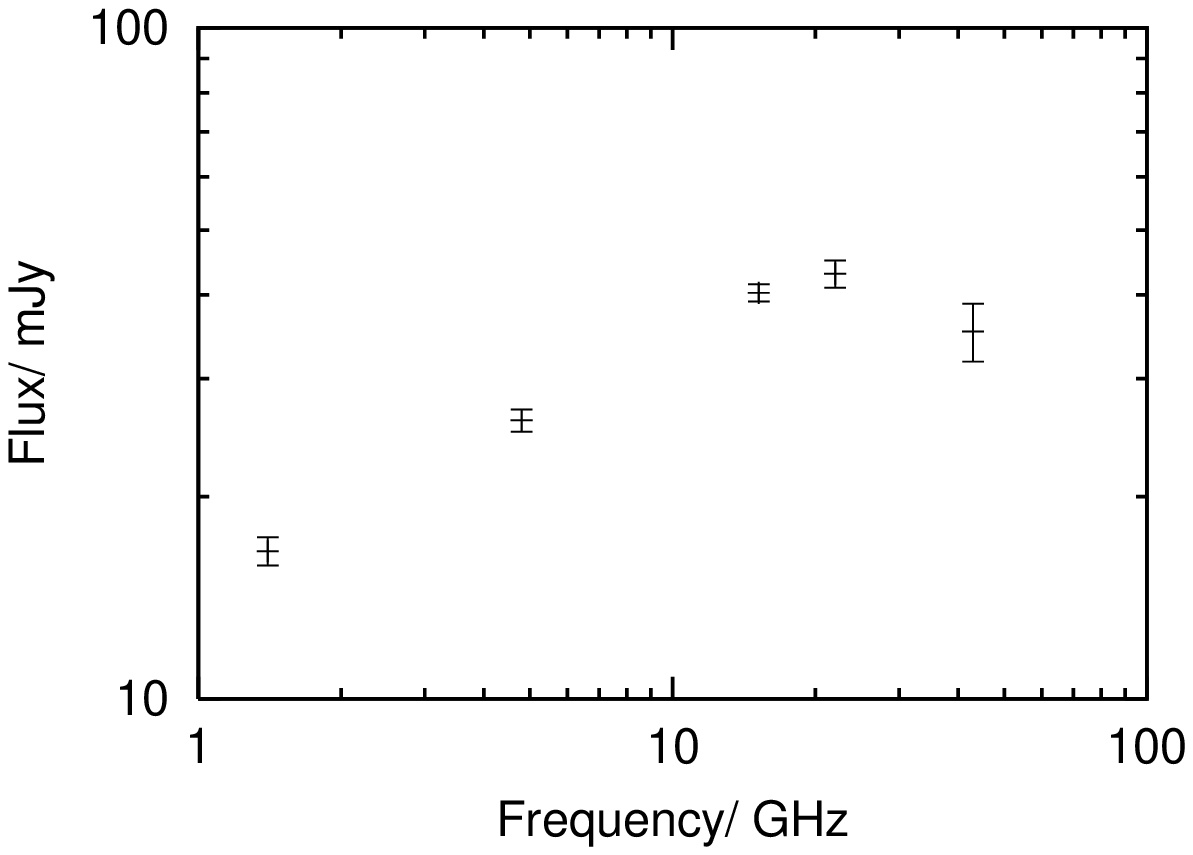,width=5cm,angle=270,clip=}\qquad
   \epsfig{figure=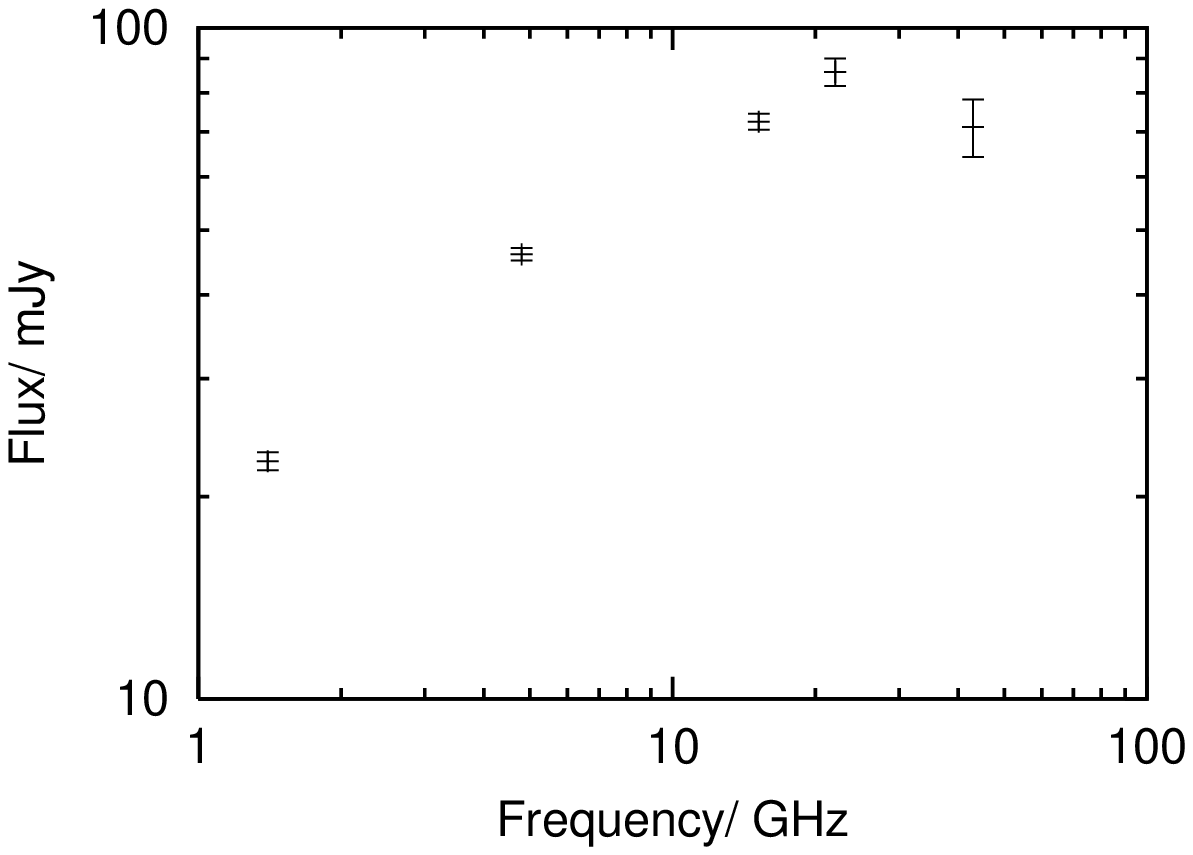,width=5cm,angle=270,clip=}}
   \caption{Radio spectra for two GPS sources peaking near 22 GHz.}
\label{GPS_spectra}
\end{figure}

Almost all previously known GPS sources have peak frequencies of order
1---5 GHz. In our sample, however, we have found a significant number
of sources which peak near the selection frequency. Of the 21 peaked
spectrum sources, 13 peak at 15 GHz or above (three are still rising
at 43 GHz). Two examples of these GPS spectra are shown in Figure
\ref{GPS_spectra}. 
If the known
trend continues---that the peak frequency increases as sources are
seen younger and younger---we speculate that these very high
frequency peakers are likely to be very young sources indeed. Efforts
are underway to develop models which track the dynamical evolution and
synchrotron spectrum to very early times.

We have found that selecting at 15 GHz is roughly twice as efficient
at picking out GPS sources than at $\sim5$ GHz, which makes it a
valuable method of creating flux-limited samples of such objects. The
9C survey, and other forthcoming cm-wave surveys, are thus useful not
only for studying the radio source population at high radio
frequencies as a whole, but also as a means of singling out
what are potentially the very youngest sources for further study.
Our optical imaging is now uncovering evidence suggesting that a
significant fraction of the host galaxies may be interacting. In the
near future we hope to quantify these processes with control samples
of both larger radiosources and non-AGN ellipticals. Plans are also in
place to select a sample at a fainter flux limit, so that we may
obtain access to sources matched in luminosity over a significant
redshift range. We are hopeful that detailed studies of this sample
and similar ones selected at high frequency will allow us both to
model precisely the earliest stages of radiosource evolution, and to
determine the exact causal connection between galaxy-galaxy
interactions and the triggering of the radiosource.
\subsubsection*{Acknowledgements}

RCB thanks PPARC for a research studentship. The Ryle Telescope is
funded by PPARC.  The VLA is operated by the National Radio Astronomy
Observatory, which is a facility of the National Science Foundation,
operated under cooperative agreement by Associated Universities, Inc.

\end{document}